\documentstyle[12pt]{article}
\begin{document}
\font\bss=cmr12 scaled\magstep 0
\title{Complex field as inflaton and quintessence}
\author{ A.V. Yurov
\small\\ Theoretical Physics Department, \small\\ Kaliningrad
State University, Russia, artyom\_yurov@mail.ru}
\date {}
\renewcommand{\abstractname}{\small Abstract}
\maketitle
\maketitle
\begin{abstract}
We investigate cosmology with complex scalar  field. It is shown that such
models can describe two stages of inflation and the oscillatory regime.
Thus we don't need both quintessence and $\Lambda$-term.
It is enough to have single complex field to obtain both inflation and present accelerated
universe. If the present accelerated expansion was created
by the complex field then universe must escape eternal acceleration.
Besides, we will show that cosmological equations with complex field
admit stationary solution  without any cosmological constant.
\end{abstract}
\thispagestyle{empty}
\medskip

\section{Introduction.}
Recently, Steinhardr and Turok suggested a new cosmological scenario called Cyclic Universe [1]. This
scenario is the development of their Ekpyrotic Universe [2]. They write "{\em The discovery of
dark energy is a complete surprise from the point-of-view of big bang and inflationary
cosmology}" [3]. Dark energy is the reason of expansion of the universe to accelerate.

Steinhardt-Turok point-of-view is too flat but there is some thing. It is easy to
formulate usual
inflationary theory to obtain recently discovered accelerated expansion
of the universe. To do it one can add a small constant term $V_0>0$ to the potential
$V(\phi)$ [4]~\footnote{Cosmology with negative potentials were considered in
[5].}.
If cosmological constant $V_0\sim 10^{-120}$ (in Planck units) then one
get the present accelaration [5], but is not terrifically because it is not clear
why should $V_0$ be so small? This approach return us to the old mystery of
vacuum energy [6].

Another way to solve the problem of accelerated expansion is a new kind of matter called
$\Lambda$-field or quintessence [7]-[8]~\footnote{Interesting alternative to quintessence is
the models like Chaplygin gas [9].}. It is clear that a new kind of
matter lead to new problems. It will be
better to obtain present inflation using the same field which was cause of inflation in
early universe. It is not clear how one can do it by-passing the problem of
$\Lambda$-term in the case of real field (vide supra) so in this work we examine
the question: can {\bf complex} field support  this role?

The main goal of this work is to show that complex field can
(as a matter of principle) lead to both inflation in early universe and present
acceleration of universe. In other words, we don't need  two different kind of
matter to understand inflation and present acceleration and we don't need
wrapped in mystery $\Lambda$-term.

We deal with qualitative investigation and we demonstrate three separate
bits of mosaic: the first inflation,  the oscillatory regime and the second inflation.
Whether the exact solution (analytical or numerical) exist where these bits
are glued in complete mosaic? We will not discuss it in this paper.

\section{Complex field.}

Complex fields in quantum and classical cosmology were introduced at first time in
[10, 11].  As we have shown  in [12], the complex field give us an inflationary
cosmology with natural exit as distinct from models with a real inflaton. In the
other hand,  the complex  field in exact soluble models (which was studied in [12])
leads to very small inflation and we supposed that a satisfactory model may contain
two scalar fields: one real  inflaton and one complex  "anti-inflaton"  i.e.
the field that triggers the end of inflation. As we shall see in the next section,
this conclusion is not valid in the case of chaotic inflation.

Let us consider Fridman-Robertson-Walker universe containing a minimally coupled complex
scalar field $\Phi(t)$ with potential $V(\mid\Phi\mid)=m^2\Phi\Phi^*$. The field
equations can be derived by the minimizing the action (in units with $8\pi G=c=1$):
$$
S=\int dt\sqrt{-g}\left(\frac{1}{2}R+g^{ik}\Phi^*_{;i}\Phi_{;k}-V(\mid\Phi\mid)\right).
$$
We choose $\Phi=\phi(t)\exp\left(i\theta(t)\right)/\sqrt{2}$ with the cyclic
variable $\theta(t)$ therefore
$$
M\equiv \phi^2{\dot\theta}=const.
$$
We call the conserved
quantity $M$ the "charge". Finally, the Einstein equations reduced to the set
$$
H^2+\frac{k}{a^2}=\frac{1}{6}\left({\dot\phi}^2+m^2\phi^2+\frac{M^2}{\phi^2}\right),\qquad
{\ddot\phi}+3H\dot\phi+m^2\phi=\frac{M^2}{\phi^2}\left(\frac{1}{\phi}-\frac{3H}{\dot\phi}\right),
\eqno(1)
$$
where $a=a(t)$ is the expansion scale factor and $H={\dot a}/a$ is the Hubble expansion
parameter. If $M=0$ then centrifugal terms are disappear and one get usual equations
for the real field $\phi$.

The centrifugal terms leads to the important difference between real  and complex
fields. For instance,  the equations (1) admit stationary solution without
any cosmological constant ($k=+1$):
$$
a=a_0=\sqrt{\frac{3}{mM}},\qquad \phi=\phi_0=\sqrt{\frac{M}{m}},\qquad H=H_0=0.
\eqno(2)
$$
(2) is the simplest solution of the system (1). In the case of general position one
can't solve these equation exactly.  Some examples of  exact soluble potentials are
contains in [12]. For example, one of these potentials has the form
$$
V=\frac{m^2\phi^2}{2}-V_0,\qquad V_0=\frac{m^2}{3}+\frac{3M^2}{4},
$$
so we have model where the potential may be become negative for $\phi<\sqrt{2V_0}/m$.
In the case of real field, cosmologies with negative potentials were investigated
in [5]. One of major conclusions is that such models enter a stage of contraction.
For the complex field we have also the case.
The great number of  confirmatory models take place in the [12].   One can show
this using the simple argumentation: let consider the cosmological equation (1) for
the arbitrary potential $V(\phi(t))\equiv V(t)$ and $k=0$. With the exception  of quantity
${\dot\phi}^2 +M^2/\phi^2$ one get the	Riccati  equation
$$
{\dot H}+3H^2=V,
$$
which can be linearized  by the substitution $H=(\log\psi)/3$. As a result, one get the
Schr\"odinger equation for the $\psi(t)=a^3(t)$,
$$
\ddot\psi=3V(t)\psi.
\eqno(3)
$$
Thus, to find the scale factor one need solve the Sch\"odinger equation (3)
with zero eigenvalue.  Let us assume that $V\to 0$ as $t\to\infty$. If $V>0$ then the
solution $\psi$ of the equation (3) is	not normalizable function. In  other
words,	$a=\psi^{1/3}\to\infty$ as $t\to\infty$.  It is not true when $V<0$ therefore
negative potential $V$ can lead to the contraction though we deal with the open
model ($k=0$).
\section{Chaotic inflation.}

One start with Linde chaotic inflation for the real field.
Let $M=0$, then the equations (1) reduces to the well known system
$$
H^2=\frac{1}{6}\left({\dot\phi}^2+m^2\phi^2\right),\qquad \ddot\phi+3H\dot\phi+m^2\phi=0.
\eqno(4)
$$
The inflation is appear when energy density is accumulated in potential, i.e.
$$
{\dot\phi}^2\ll m^2\phi^2,\qquad \mid\ddot\phi\mid\ll m^2\mid\phi\mid.
\eqno(5)
$$
Using (4), (5) one get $H^2=m^2\phi^2/6$ and $3H\dot\phi=-m^2\phi$ therefore
$$
\phi(t)=\phi_0-\sqrt{\frac{2}{3}} mt.
\eqno(6)
$$
The conditions (5) are valid if  $\phi_0\gg 1$ and $t\ll \sqrt{3/2}\phi_0/m\equiv t_e$,
where $t_e$ is the time of exit from inflation. During inflation ($t<t_e$) one may
suppose that $\phi\sim\phi_0$  so $H=m\phi_0/\sqrt{6}$=const. In other words, we get
the de Sitter-like regime. The number of e-foldings  at time $t_e$  is
$\alpha(t_e)=\phi_0^2/2$ therefore the condition $\phi_0\gg 1$ lead to good inflation.
For example, if $\phi_0=10$ then $\alpha(t_e)\sim 50$. We will hold $\phi_e=\phi(t_e)=1$.

All that is well known Linde chaotic inflation for the real inflaton. Let consider
the equation for complex field (1). These equation are differ from  the (4) because
the system (1) has additional "centrifugal terms".  The Linde inflation will be valid
if "centrifugal terms" will be small, i.e.
$$
\frac{M}{m}\ll\phi^2,\qquad \frac{M^2}{m^2}\ll \frac{\mid\dot\phi\mid\phi^3}{3H}.
$$
The first condition will be true if $M\ll m$. It is strong inequality
guaranteeing that $M^2/\phi^3\ll m^2\phi$ right up to the end of inflation.  The second
condition can be written  as $M/m\ll \sqrt{2}\phi_0/\sqrt{3}$, i.e. it is coincide
with the first, practically.

Thus, usual Linde chaotic inflation can take place for the complex field
if  $M/m\ll \phi_0$. Nevertheless, we should use more strong condition $M/m\ll 1$ which
we'll  need later. In this regime one can neglet   "centrifugal terms" and complex field
is indistinguishable from the real one.
\section{The oscillatory regime.}

One should continue the analogy between real and complex field for the
oscillatory regime.  We start from the real field. If the $\phi$ oscillates
near $\phi=0$ with frequency $m\gg H$ then the second equation of the system
(4) can be written as
$$
\ddot\phi+m^2\phi=0.
\eqno(7)
$$
One find the solution of (7) in the form $\phi=\Phi(t)\sin mt$, where $\Phi(t)$ is
the slow variable function.  Neglecting  $\dot\Phi$ one calculate the pressure:
$$
p=\frac{1}{2}{\dot\phi }^2-\frac{m^2\phi^2}{2}=\frac{m^2\Phi^2}{2}\cos 2mt,
$$
so taking an average over many oscillations one get $<p>=0$. Thus we have the
dust equation of state. In this case the solution of Einstein  equations is
well known: $a\sim t^{2/3}$ and $H=2/3t$. Substituting $H$ into the first equation
of (4) one get
$$
\phi=\frac{2\sqrt{2}}{3mt}\sin mt.
\eqno(8)
$$

It's clear that  (8) can't describe oscillations of complex field (even
approximately). This
because $\phi=0$ at $t_{_N}=\pi N/m$ (and $\dot\phi=0$ at $t=t_{_N}+\pi/2m$)
therefore centrifugal terms are not small close to these points.
Maybe we can use this regime during the time when $\phi\ne 0$ and $\dot\phi\ne 0$?
It is not right because during the time the number of oscillations less than one.
Thus one need
another oscillatory regime. Fortunately, it is easy to find  one.

Now let us assume that the real field  $\phi$ oscillates near $\phi=A(t)\ne 0$:
$$
\phi=A(t)+\Phi(t)\sin mt.
\eqno(9)
$$
Calculating the pressure  one get
$$
p=\frac{1}{2}\left({\dot A}^2-m^2A^2+(m\Phi+2{\dot A})m\Phi\cos mt-2m^2A\Phi\sin mt\right),
$$
therefore
$$
<p>=\frac{1}{2}\left({\dot A}^2-m^2A^2\right).
$$
If one need to obtain $a\sim t^{2/3}$ then $<p>=0$ so $A(t)=A_0 e^{k mt}$ with
$k=\pm 1$. Substituting the expression for the $A(t)$ into the (9) and (9) into the first
equation (4) we have
$$
\phi=A_0e^{k mt}+\sqrt{\frac{8}{9m^2t^2}-2A_0^2 e^{2k mt}} \sin mt.
\eqno(10)
$$
(10) is generalization of  (8). If $A_0=0$ then
(10)=(8)~\footnote{Note, the condition $<p>=0$ is not  necessary.  One can use some
another condition.}.

Let $z=mt$, $\theta=A_0z e^{k z}$. There are two condition which lead to
good behavior of (10): the radicand must be positive and $\phi\ne 0$.
Both of them can be written as compact inequality,
$$
\frac{2\sqrt{2}}{3}<A_0ze^{k z}<\frac{2}{\sqrt{3}}.
\eqno(11)
$$
It is clear that the oscillatory regime work during finite time whereupon one of
inequalities (11)  will be break.
There are two different cases when $k=1$ (i) and $k=-1$ (ii).
If $k=-1$ then one get two cases (iia) and (iib).
The case (iia) take place when $2\sqrt{3}>A_0/e>2\sqrt{2}/3$ while (iib) take place
when $A_0/e>2\sqrt{3}$. The time of this regime can be sufficiently large
to create enough number of  pairs.

Now let us consider  the same regime for the complex field. We need that
"centrifugal terms" will be small during regime of oscillations  (10).
It is possible if  the reason of exit from this regime is the violation
of second member of two-sided inequality (11). In this case $\phi$ and $\dot\phi$ will
be nonvanishing function right up to intersection of curves $\theta$ and
$2/\sqrt{3}$ therefore the "centrifugal terms" will be finite and we can
choose $M$ so small to suppress these terms. Suitable situations are (i) and (iib).

We need just these cases to obtain small "centrifugal terms" both  during and
{\bf just on completion} of the oscillatory  regime. If $\phi$ is the decreasing
function of $t$ then we have the point of time $t_*$ such that "centrifugal terms" are
the first order of smallness. Taking into account these terms via the perturbation
theory	we have a new dynamics which is characteristic precisely of the complex field.
As we shall see, that will do to obtain new inflation.
\section{Second inflation.}

Let us consider the system (1). We suppose that all energy density is dominated
by the effective potential $U=U(\phi)$,
$$
\frac{1}{2}{\dot\phi}^2\ll U,\qquad  \mid\ddot\phi\mid\ll\mid U'\mid,
\qquad U\equiv \frac{m^2\phi^2}{2}+\frac{M^2}{2\phi^2}.
\eqno(12)
$$
So $H=\sqrt{U/3}$. After simple calculations we get two possible expressions for
the $\dot\phi$:
$$
{\dot\phi}_{\pm}=\frac{1}{2\phi\sqrt{U}}\left(-U'\phi\pm \sqrt{(U'\phi)^2-12M^2U}\right).
$$

Let $\phi_0$ is the  value of $\phi$ at which  inflation begins. Besides, we'll use the
condition $\phi\gg \sqrt{M/m}$ which was obtained above. Using Taylor's series
($M$ is the small parameter) we get
$$
{\dot\phi}_+\sim -\sqrt{\frac{3}{2}}\frac{M^2}{m\phi^2},\qquad
{\dot\phi}_-\sim -\sqrt{\frac{2}{3}}m.
$$
The solution $\phi_-$ is the usual Linde inflation (6) whereas
the $\phi_+$ is new solution which is characteristic precisely of the complex field.
After integration we have
$$
\phi_+=\left(\phi_{0,+}^3-\sqrt{\frac{3}{2}}\frac{3M^2t}{m}\right)^{1/3}
\eqno(13)
$$
The condition (12) can be written as
$$
\phi\gg\left(\sqrt{\frac{3}{2}}\frac{M^2}{m^2}\right)^{1/3}
$$
and it is jointly satisfiable with $\phi\gg\sqrt{M/m}$. The number of
e-foldings during the new inflation ($\alpha_+(t)$) and during Linde
inflation ($\alpha_-(t)$) are
$$
\alpha_+(t)=\frac{m^2}{12 M^2}\left(\phi_{0,+}^4-\phi^4(t)\right),\qquad
\alpha_-(t)=\frac{1}{4}\left(\phi_{0,-}^2-\phi^2(t)\right).
$$
Thus, to the end of inflation the number of e-foldings will be
$$
\alpha_+(t_e^{(+)})\sim \frac{m^2\phi_{0,+}^4}{12 M^2},
\qquad t_e^{(+)}=\frac{\sqrt{2} m\phi_{0,+}^3}{\sqrt{27} M^2},
$$
while for the Linde inflation we have $t_e^{(-)}=\sqrt{3}\phi_{0,-}/2m$.

During the time $t\ll t_e^{(+)}$ we have $\phi\sim\phi_{0,+}$
and $H=\sqrt{U(\phi_{0,+})/3}=$const, so we have de Sitter-like regim.
Thus, if the $M$ is small the equations (1) admits  two inflations.
The first (6)  is usual Linde chaotic inflation which took place in the early universe.
The second (13) is new and take place only for the universe with complex scalar field.
We get the designing question: it is possible to  equate this regime with recently
discovered accelerated expansion of the universe? It will be very nicely to obtain the
positive answer. This because such explanation has two advantages:
\newline
\newline
1. Both the inflation in the early universe and the present accelerated
expansion of universe are sequent of existence and dynamics of single object - complex
scalar field. We don't need "quintessence".
\newline
2. We don't need cosmological constant  so we have not problems with it's anomalous small value.
\newline
\newline
The time of the first inflation is $t_e^{(-)}\sim 10^{-35}$ s. The time
$t^{(+)}$ is counts off from the time when the second inflation began.
We see that this time is finite,  i.e. if the present accelerated expansion was created
by the complex field then universe must escape eternal acceleration. The time of the
second inflation $t_e^{(+)}$ must be much more long than $t_e^{(-)}$ if the
second inflation is the present acceleration. We choose $\phi_{0,+}=1$,
therefore $t_e^{(+)}/t_e^{(-)}=4m^2/9\phi_{0,-}M^2$. This ratio  will be
larger then unit if
$$
\frac{M}{m}\ll \frac{2}{3\sqrt{\phi_{0,-}}}.
\eqno(14)
$$
(14) is strong inequality and it lay on $M$ very hard restriction
(in  comparison with the inequality $M/m\ll 1$). In units with $\hbar=c=1$ we have
$M\ll 0.42\times 10^{27}$ GeV${}^3$ and if $t_e^{(+)}\sim 10^{10}$ years then
$M\sim 1900\phi_{0,+}$ GeV${}^3$ if
$\phi_{0,-}=10^{25}$ GeV and $m=10^{12}$ GeV.
\section{Conclusion.}
Thus, complex field can be both inflaton and quintessence.  The aim of this work is to show that a new way to solve well known cosmological problems is exist in principle. The next step is the search of exact solution (numerical or analytical)  where  these bits are glued in complete mosaic:
\newline
\newline
the first inflation$\to $ the oscillatory regime $\to$ the second inflation.
\newline
\newline
It is possible that such solutions are not exist. That would be a pity because this way to solve both problems of the inflation (with exit)  and the present accelerated expansion of universe is looking very fine.
$$
{}
$$
{\em Acknowledgements.}
This work was supported by the Grant of Education Department of the Russian Federation,
 No. E00-3.1-383.
$$
{}
$$
\centerline{\bf References}
\noindent
\begin{enumerate}
\item P.J. Steinhardt and N. Turok,\rm\,"A Cyclic Model of the Universe",
hep-th/0111030.
\item J. Khouru, B.A. Ovrut, P.J. Steinhardt and N. Turok,\rm\, Phys. Rev. D
{\bf 64}, 123522 (2001).
\item P.J. Steinhardt and N. Turok,\rm\,
astro-ph/0204479.
\item A. Linde, \rm\, hep-th/0205259.
\item G. Felder, A. Frolov, L. Kofman and A. Linde,\rm\,
hep-th/0202017.
\item A.D. Dolgov, \rm\, hep-ph/0203245.
\item R.R. Caldwell, R. Dave and P.J. Steinhardt,\rm\, Phys. Rev. Lett.
{\bf 80}, 1528 (1998).
\item V. Sahni and A. Starobinsky,\rm\, Int. J. Mod. Phys. D {\bf 9}, 373 (2000).
\item A. Kamenshchik, U. Moschella and V. Pasquier,\rm\, gr-qc/0103004.
\item A.Yu. Kamenshchik, I.M. Khalatnikov and A.V. Toropensky,\rm\, gr-qc/9508034 v1.
\item A.Yu. Kamenshchik, I.M. Khalatnikov and A.V. Toropensky,\rm\, gr-qc/9801039 v1.
\item A.V. Yurov,\rm\, Class. Quantum Grav., {\bf 18}, 3753 (2001).

\end{enumerate}

\end{document}